# Assessing Age Assurance Technologies: Effectiveness, Side-Effects, and Acceptance

Wouter Lueks (CISPA Helmholtz Center for Information Security)
Stephan Dreyer (Leibniz Institute for Media Research | Hans-Bredow-Institut)
Hannes Federrath (University of Hamburg)
Judith Simon (University of Hamburg)

## Abstract

In this paper, we provide an overview and evaluation of different types of age assurance technologies (AAT). We describe and analyse 1) different *approaches to age assurance* online (age verification, age estimation, age inference, and parental control and consent), as well as 2) different *age assurance architectures* (online, offline device-based, offline credential-based), and assess their various combinations with regards to their respective a) effectiveness, b) side effects, and c) acceptance. We then discuss general limitations of AAT's effectiveness stemming from the possibility of circumvention and outline the most important side effects, in particular regarding privacy and anonymity of all users; bias, discrimination, and exclusion; as well as censorship and related concerns. We conclude our analyses by offering some recommendations on which types of AAT are better or less suited to protect minors online. Guiding our assessment is a weighing of effectiveness against side effects, resulting in a graduated *hierarchy of acceptable AAT mechanisms*.

# 1 Introduction

This discussion paper is meant to provide an overview and assessment of the different types of age assurance technologies (AAT) developed or already deployed to protect children and young adults from harmful content online. We describe different *approaches to age assurance* online (age verification, age estimation, age inference, and parental control and consent), different *age assurance architectures* (online, offline-device-based, offline credential-based) and assess their various combinations in regards to their respective a) effectiveness, b) side effects, and c) acceptance. Importantly, any measure proposed must account for and balance the three core desiderata of children rights in the digital environment, namely protection, participation, and empowerment. Based upon our analyses, we offer recommendations on which types of AAT are better or less suited to protect minors online. Guiding our assessment is a weighing of effectiveness against side effects, resulting in a graduated *hierarchy of AAT mechanisms*. We hope that our portrayal of the state of the art can serve as a basis for a more nuanced deliberation and decision making about the pros and cons of different solutions proposed in this quickly evolving field.

## 1.1 Background

In recent months, we have witnessed a surge in discussions concerning potential negative effects of social media on children's and young adults' mental well-being, leading not only to academic, but also public and political debates whether there is a need for action and if so, which type or combination of actions may be best suited to address risks and mitigate harms.

Governments in various different countries have either already proposed a ban of social media for children under certain ages or are currently discussing similar measures. An early example, which received much attention, is the so-called social media delay for children under the age of 16 in Australia which went into force on December 10, 2025, following an intense debate as to whether this measure is proportionate on the one hand, and effective on the other. Greece also introduced a social media ban for youngsters under the age of 15 starting in October 2025, and France, Portugal, Austria and Denmark plan to do so in 2026. The UK mandated age verification by the Online Safety Act, and requires websites with age-restricted content (especially pornography) to confirm users are at least 18 years old, while Germany has a comparable legal obligation in place since 2002.

Reactions to the implementation of such age restrictions and the social media bans in particular have been mixed. While some observers have applauded the initiatives as a much needed intervention, which should be replicated elsewhere, others have raised various issues against this measure. Most of these critical voices were not denying the potential harms posed by social media, but either 1) stressed the important role social media plays for communicative and informational practices of young people or 2) pointed to the negative side effects which measures required to keep minors off of social media could have.

Arguments regarding the positive role of social media highlighted that digital technologies in general – and social media in particular – are so interwoven into contemporary live for adults as much as for children, that simply blocking all access to social media might severely constrain the informational, communicative and social needs of children, thereby also infringing children's rights to information, education and freedom of expression. Accordingly, any assessment of and recommendation regarding instruments which aim at the protection of children and teenagers online must weigh the harms potentially resulting from unrestricted access with the potential harms resulting from restricting access itself. This points to one central ethical dilemma: measures to protect minors from harms online may infringe their rights and those of others, and thus cause different harms.

Arguments pointing to potential negative side effects, in contrast, did not focus on the benefits and harms of social media, but rather at the possibilities and limitations of age assurance technologies, i.e. those technologies which are required to reliably distinguish between adult users and minors. These systems need to meet at least two central, yet often opposing demands: they need to 1) be effective, i.e. their deployment should disable or at least substantially harden forbidden access as well as 2) have no side effects which are disproportionate. Such side effects for AAT most notably concern undue invasions of privacy, security concerns, digital exclusion or undue costs of digital sovereignty, e.g. through reliance on platforms, to name only the most salient concerns. Moreover, it would 3) need to be accepted by users as low acceptance would hinder uptake or set incentives for circumventions.

Conceptually and technically, limiting the exposure of children and youth to content considered harmful online requires effective means on two ends: a classification of users on the one side and a classification of content on the other, combined with mechanisms of matching the two. In this paper, we focus only on the former process, the reliable classification of users into age cohorts. As they are a prerequisite for providing age-appropriate features and content and for many measures of access restrictions, it is of pivotal relevance to assess the possibilities, limitations and implications of different types of age assurance technologies (AAT). The classification of content according to age-appropriateness is a complementary, yet entirely different process, entailing different regulatory and technological means (cf. for instance Winder et al. 2023).

## 1.2 Overview

This discussion paper is structured as follows. In **Section 2**, we briefly outline the relevant **regulatory context** for age assurance technologies. Apart from German and European law, we also portray the UK Online Safety Act. As it claims extraterritorial application, every service that is accessible in the UK has to adhere to it and may thus be of relevance for European and international platforms.

**Section 3** provides a systematic overview and assessment of the different **approaches to age assurance**, namely: **age verification, age estimation, age inference, and parental control and consent**. For each approach, we 1) offer a brief explanation of their technological foundations, 2) assess their respective strengths and risks, and 3) provide first empirical indications on their acceptance based upon survey data. The section ends with preliminary conclusions regarding assurance levels, side effects, acceptance, and applicability of all approaches.

**Section 4** provides an overview over three different **age assurance architecture**s, i.e. three different ways in which results of age assurance can be transferred to service providers, such as social media platforms or porn websites. We distinguish between *online assurance* and two forms of offline assurance, *device-based* and *credential-based offline age assurance*. We first discuss the compatibility between these three architectures and the four approaches presented in Section 3. We then present and evaluate the three architectures in more detail, focusing both on their level of assurance regarding effectiveness and their side effects. Section 4 ends with some preliminary conclusions regarding the different combinations of age assurance approaches and architectures.

After having assessed the details of the different approaches and architectures, **Section 5** offers a discussion of the **general limitations and side effects** of measures for age assurance. We first discuss general limits to the *effectiveness* of proposed measures resulting from various means for *circumvention*. We then outline the most important side effects of age assurance technologies, namely infringements of *privacy and anonymity*, issues of *bias, discrimination and exclusion*, as well as *censorship* and the risk of *closing of the internet* as a whole *for all users*.

In **Section 6**, we end with some **conclusions** as to whether and if so which types of age assurance technologies we consider acceptable and appropriate to protect minors from harmful services online. Readers interested only in the results of our analysis may jump directly to this section and refer to earlier ones for the substantiation of our conclusions.

# 2 Regulatory Context

Debates on age verification and age assurance techniques are not taking place in a lawless space; actually, there is a complex legal framework governing online content and services that is also legally shaping current discussions. At the EU level, three legislative acts touch age assessment use cases: the Digital Services Act (DSA)[1], the Audiovisual Media Services Directive (AVMSD)[2] and the General Data Protection Regulation (GDPR)[3]. In parallel, prominent approaches concerning age assessment obligations on the level of national legislation can be found in the United Kingdom's Online Safety Act 2023 (OSA)[4] and Germany's Jugendmedienschutz-Staatsvertrag (JMStV)[5]. These legal frameworks define either duties or at least permissible options for service providers when deploying age assessment systems.

## 2.1 Legal Framework for Age Assessment in the EU

Across all EU Member States, the DSA establishes horizontal obligations for so-called intermediary services, with specific provisions for online platforms: Providers of online platforms "accessible to minors" must implement appropriate and proportionate measures to ensure a high level of privacy, safety and security for minors (Art. 28(1) DSA). For very large online platforms and search engines, age assurance is expressly listed as a possible systemic risk-mitigation measure, particularly in relation to risks to minors' mental health and exposure to harmful content (Art. 35 DSA). At the same time, the DSA clarifies that platforms are not obliged to process additional personal data to determine users' age (Art. 28(3) DSA), and prohibits targeted advertising based on profiling where the provider "knows with reasonable certainty" that the recipient is a minor. The recently published Commission guidelines[6] under Art. 28 (1) DSA interpret these provisions as encouraging "effective age assurance methods" for high-risk contexts, e.g. pornography, gambling, and certain social media features. The guidelines introduce five distinct categories when assessing age assessment technologies in the context of the platform providers' duties of care: First, assessment focuses on how precisely an age verification or age estimation method determines whether a user is above or below a defined age threshold or within a specific age range (accuracy). The reliability category concerns the consistent functioning of the method under real-world conditions rather than idealized laboratory settings. A reliable method must be continuously available and operational across diverse contexts of use; providers must verify that all data sources used in the age assurance process are trustworthy. Robustness refers to the resistance of a method to circumvention, particularly by minors. A method that can be easily bypassed is deemed ineffective. The assessment of robustness must be contextual and age-specific, taking into account the capabilities of the minors addressed. Providers must also ensure that the method incorporates appropriate technical and organizational safeguards to protect the integrity and security of age-related data in line with the state of the art. Non-intrusiveness evaluates the extent to which an age assurance method interferes with users' fundamental rights, notably privacy, data protection, and freedom of expression. Only age-related attributes that are strictly necessary for the specific purpose may be processed, in accordance with Article 28(3) of the Digital Services Act and the statement of the European Data Protection Board (EDPB, 2025). Where multiple methods provide comparable effectiveness, the least intrusive option must be chosen. The assessment regarding non-discrimination examines whether an age assurance method disadvantages or excludes certain users or groups of users. Providers must ensure that the method is accessible and appropriate for all minors, irrespective of disability, language, ethnic origin, gender, religion, or minority status. In practice, this positions age assessment measures with regard to relevant online platforms as recommended – maybe even politically expected – tools for protection, but not as an outright legal obligation.

---

[1] https://eur-lex.europa.eu/legal-content/EN/TXT/?uri=OJ:C_202505519
[2] https://eur-lex.europa.eu/legal-content/EN/TXT/?uri=CELEX%3A02010L0013-20250208
[3] https://eur-lex.europa.eu/legal-content/EN/TXT/?uri=CELEX%3A02016R0679-20160504
[4] https://www.legislation.gov.uk/ukpga/2023/50
[5] https://www.kjm-online.de/fileadmin/user_upload/Rechtsgrundlagen/Gesetze_Staatsvertraege/JMStV/JMStV_english_version.pdf

The same goes for the legal framework set out by the Audiovisual Media Services Directive (AVMSD): For broadcast and on-demand services, Member States must ensure that content which may impair the physical, mental or moral development of minors is made available in such a way that minors will not normally hear or see it (Art.

---

[6] https://eur-lex.europa.eu/eli/C/2025/5519/oj

6a AVMSD). Acceptable measures explicitly include age verification systems, alongside, inter alia, watersheds. With regards to video-sharing platform services, providers must take appropriate measures to protect minors from user-generated videos and commercial communications that may impair their development, and those measures might include age verification systems among others (Art. 28b AVMSD). In contrast to the fully harmonising and directly applicable DSA provisions, necessary national implementations of the AVMSD differ in detail but the AVMSD effectively normalises some form of age gating or age assurance for harmful audiovisual content, particularly pornography and extreme violence, across the EU.

Lastly, the GDPR as the central legal framework for data protection does not regulate content but the processing of personal data. Article 8 GDPR introduces the notion of an "age of digital consent" (13-16, depending on Member State law) and requires controllers to make "reasonable efforts" to verify that a person who gives consent towards any digital service is above this threshold, or otherwise that parental consent is given. The European Data Protection Board guidance stresses that verification methods must be proportionate to respective risks of the data processing in each case and that they have to comply with data protection principles such as data minimisation and purpose limitation. As a result, extensive document-based verification for each and every online service will typically be disproportionate for low-risk processing, whereas higher risk use cases (e.g. access to adult content) can justify more robust checks, as long as privacy-enhancing techniques are used. In combination with the DSA and AVMSD, the GDPR thus pushes designers of age assurance systems towards privacy-preserving solutions like pseudonymous age tokens, local or one-shot age estimation rather than centralised databases of identity documents. An issue yet unsolved is the aspect that age assessment services will usually fall under Art. 8 (1) DSA themselves, resulting in a vicious circle, i.e. the provider has to receive information about the age of the person to be age-assessed to decide whether parental consent is to be sought.

## 2.2 National Legal Frameworks in the UK and Germany

In the United Kingdom, the Online Safety Act from 2023 introduces a more prescriptive framework: Services that publish or allow pornographic content and are accessible from the UK must ensure that persons under 18 "are not normally able to encounter" such content, and Ofcom's guidance (Ofcom, 2023) makes clear that "highly effective" age verification is required for this purpose. Simple self declaration ("I am over 18") is explicitly insufficient. Acceptable options include third-party digital ID checks, credit or payment-card based checks, AI-based facial age estimation, mobile network age checks and privacy-preserving digital identity wallets. Beyond pornography, user-to-user and search services "likely to be accessed by children" must undertake risk assessments and use proportionate measures including age assurance to prevent children's exposure to content designated as "primary priority content" or to provide an age-adequate access to content that is deemed "priority content". In parallel, the Age Appropriate Design Code (ICO, 2022) under UK data protection law requires services likely to be accessed by children either to treat all users as children or to deploy reliable mechanisms to distinguish children from adults and adjust data practices and features accordingly. This way, the UK regime shifts age assessment for some sectors, particularly pornography, from a policy choice to a hard legal duty, with strong regulatory expectations for age assurance on mainstream platforms as well.

Compared to the UK case, Germany's JMStV represents a legal approach to content-specific youth protection that stands for more than 20 years. It distinguishes between content that is "harmful to minors" and content that is merely "impairing minors' development". For the former category that includes hardcore pornography or excessive violent content, providers may only offer the content in closed user groups that are effectively inaccessible to minors. In practice, this requires state- or self-regulatory body-approved age verification systems combining reliable identification with secure authentication on each access. For impairing content, e.g., material rated 16+, providers must deploy "adequate means" to ensure that younger minors will not normally encounter it, such as watershed scheduling, electronic age labels combined with certified filter software or parental controls, or lighter forms of age assessment (e.g. providing the identity card number). A current reform of the JMStV has introduced duties for providers of operating systems and app markets, requiring them to integrate parental controls on the OS and app distribution levels. This way, the German legal framework currently contains a hard legal duty for providers of 18+ content to implement strict age verification hurdles, and for providers of 16+ content to at least provide measures to check the age of a user at least in a "plausible" way. For both scenarios, the competent KJM (Kommission für Jugendmedienschutz) and the approved co-regulatory bodies have provided more detailed requirements for adequate age assessment solutions (KJM, 2022).

Moreover, according to § 19 (2) of the German Telecommunications Digital Services Data Protection Act (Telekommunikation-Digitale-Dienste-Datenschutz-Gesetz, TDDDG), "providers of digital services must enable the use of digital services and their payment anonymously or under a pseudonym, insofar as this is technically feasible and reasonable. Users of digital services must be informed of this option." As many of the age verification and age assurance techniques require age-verified user accounts, even if the digital service does not require to register (i.e. browsing web pages, watching public and freely available TV streaming services), the (technical) enforcement to register and create user accounts may be in conflict with this TDDDG provision.

## 2.3 Age Assessment as an EU Option and a National Duty

Against the background of the multi-level governance frameworks when it comes to the protection of minors online, structural tensions and uncertainties can be observed: First, there is regulatory fragmentation, with EU law providing principle-based duties for online platforms and video-sharing platform providers but no strict age assessment obligation, while Germany and the UK have adopted comparatively strict, prescriptive requirements in certain sectors. Content providers operating cross-border must therefore meet different thresholds of robustness and different definitions of "effective" age assurance. Second, there is a yet unresolved privacy versus safety trade-off: The DSA and the AVMSD emphasise that platforms are not obliged to collect additional personal data solely to identify minors, yet robust exclusion of minors from high-risk content almost inevitably requires some form of identity- or attribute-based verification. This raises questions about acceptable accuracy, data minimisation and the long-term implications for anonymity. Third, the scope of protection is uneven across content types: While audiovisual media services providing adult content are usually tightly regulated, text-based, image-based or interactive services offerings with content for all age groups may fall into less severe provisions. All online platforms offering user-generated content are captured via the DSA's general platform duties, resulting in a large leeway of decision-making when it comes to selecting the appropriate platform measures, including age assessment. Finally, discrepancies in age thresholds and concepts – GDPR's "digital consent" age and national youth protection age ratings – complicate harmonised implementation of age assurance technologies in practice.

The current legal frameworks on EU and state level do not prescribe a single model of age verification, but they increasingly condition service designs in the direction of technical solutions to protect minors. While EU laws incentivise risk-based, privacy-respecting age assurance, the two examples for national legal frameworks (UK and Germany) mandate robust age verification in specific high-risk sectors and expect active age differentiation across relevant mainstream services. German law exemplifies a structured, content-tiered model anchored in certified age verification systems for content providers. Any deployment of age verification techniques must therefore be read against this multi-layered regulatory backdrop, which both enables and constrains technological choices, but always binds the technical implementation to the principle of proportionality.

# 3 Age Assurance Approaches: A Systematic Overview

Following the classification outlined by the Australian Pilot on Age Assurance Technologies (AAT, 2025), we distinguish several high-level approaches to age assurance:

- **Age Verification:** The age is verified based on authoritative documents. We include here claims by parties such as banks, governments, etc. that have existing relationships with users.
- **Age Estimation:** The age is estimated based on biometric features, such as voice, face, body or movement.
- **Age Inference:** The age is inferred based on behavioral traces of a user.
- **Parental Control and Consent**: Parents have to approve access (control: a priori, consent: after access attempt has been made).

Further approaches exist in the context of protections of minors: for **self-declaration**, where a user self-declares their age ("I am above 18") as well as for **time-limited access** to content, i.e. movies being available only after 22:00 (so-called watersheds), no age assurance is enforced. We mention these approaches for the sake of completeness, but will in the following focus only on Age Inference, Age Estimation, Age Verification and Parental Control and Consent as neither self-declaration nor time-limited access to content require any technical identification of minors.

## 3.1. Empirical Research on Public Attitudes Towards Age Assurance Measures in General

As online age assurance measures are increasingly discussed publicly, empirical research has also started to investigate public perceptions of such measures. As will be detailed below, different approaches to age assurance vary both in *effectiveness* but also in *intrusiveness.* Empirical studies mirror this and find that public attitudes toward age assurance measures are indeed shaped by perceptions of their effectiveness in keeping minors safe on the one side, balanced against concerns about privacy, data security, and personal freedom on the other side. The surveys and studies mentioned below generally show broad support for the concept of age-based restrictions online, yet significant ambivalence about the methods used to implement them. In general, people want to protect children online, but they also worry about how age assessment technologies might impact privacy and civil liberties.

In principle, the public strongly agrees that adult-only websites and services should verify users' ages. 78% of Australian adults surveyed in 2021 supported a government-mandated age verification system to restrict under-18 access to online pornography (eSafety Commissioner, 2021). Similarly in the UK, 78% of adults polled in 2023 agreed that porn sites should require age checks, with only 5% opposed (Sleigh, 2023). There is a broad consensus – often around 70-80% support – that minors should be kept out of adult content through strict verification measures. Even for social media accounts, where the issue is more nuanced, about 71% of U.S. adults favor requiring users to verify their age upon signup (Pew, 2023); support for these requirements spans different demographics in surveys. Notably, parents tend to be especially supportive of age checks to protect children online (Pew, 2023).

When interpreting the results of empirical studies on acceptance a caveat may be worth mentioning: As acceptance of technologies is essential for their uptake, it is important to empirically inquire the stance of the public towards such technologies. However, it is important to note that what people report in such surveys does not have to conform to how they will act. This can be the result of various factors and biases, such as social desirability, but also a lack of knowledge about these technologies. Thus, empirical research investigating trust in and acceptance of technologies may conflate a *stance towards* such technologies with actual usage *practice*s.

Moreover, from a normative perspective users should trust or accept systems if and only if they are indeed trustworthy – and not if they merely appear to be so. Put differently: what needs to be avoided are users trusting untrustworthy systems which harm them. To conclude, survey data on trust and acceptance of technologies must be interpreted with a grain of salt as they 1) empirically may fail to distinguish a theoretical stance towards technologies from actual (future) use practices and 2) normatively fail to distinguish between appropriate and inappropriate trust in or acceptance of technologies.

## 3.2 Age Verification

### 3.2.1 Age Verification: Technological Foundations

Age verification methods rely on official sources of information to determine the age of users. We distinguish two types of age verification: (1) direct document-based verification where users provide official documents to verify their age; and (2) relation-based verification, where users rely on other parties such as a government entity to attest to their age. While both rely on official documents to bootstrap the verification, the processes differ, each having distinct advantages and disadvantages.

*Direct Document-based Verification.* When using documents, users present an official document to a verifying party. As we shall see later, this verifying party could be the service provider itself, or a third-party age assurance service. Typical examples of such a document are ID cards, driver's licenses, passports, credit card details etc. that directly attest the user's age.

The content of these documents can be transferred to the verifying party in different ways: The most common is to take a picture of the document using the camera (or webcam) on the user's device. Other solutions rely on using digital versions of ID documents where the documents are stored on the user's device or presented to the user's device and read via, e.g. an RFID chip. These latter options enable stronger guarantees about the correctness of the data obtained from the document and do not differ substantially from use cases which require strong identification,[7] such as banking apps. Thus, high levels of assurance are possible.

Often, reading the documents is combined with another mechanism to ensure that the owner of the document is *present.* This can entail a face-to-face meeting, taking a selfie, a video of one's face, or even a combination of a video and the showing of the document.

*Relation-based verification.* This type of indirect document-based verification does not rely directly on documents, but instead on existing relationships between users and 3rd parties that already *know* the user, i.e. the user has identified him/herself prior to this 3rd party, called an *identity provider (IdP)*, usually with an official document. Examples of such IdPs include government agencies, health insurers, banks, mobile service providers, etc.

The age verification process proceeds as follows (see for more details the age assurance architectures below): a user logs in to their IdP using an IdP-specific process; for example their authenticator app to log in to their bank. The IdP now has identified the user, can look up their age (and other relevant information), and subsequently transmits this information to the verifying party.

---

[7] See for example the UK ETA app guide at https://www.gov.uk/guidance/using-the-uk-eta-app for electronic travel authorisation (ETA) to visit the UK.

### 3.2.2 Age Verification: Analysis

*Effectiveness & Assurance Level:* In both document-based and relation-based age verification, the assurance level provided is high. For document-based verification the high-level of assurance directly follows from the strong assurance provided by the underlying documents. In the case of relation-based verification, the high level of assurance follows from the identification processes followed by the IdP. When the IdPs themselves base their identification processes on official documents (as for example banks and governments do), the resulting assurance is again high.

*Risk-Benefit-Analysis (general).* The advantage of age verification is independent of whether verification happens based on documents or preexisting relationships.

- Age verification inherits the high level of assurance from the underlying official identity documents.

The following risks apply independent of whether verification happens based on documents or preexisting relationships.

- *Exclusion Risk.* In both cases, users require access to appropriate official documents or a preexisting relationship with an identity provider for the purpose of verifying their age. Thus, there is the risk of *exclusion.* As we discuss in more detail later, it is not clear that adults will always be able to meet this bar. Worse, young people, especially those under the age of 16, might not have access to official documentation, such as an ID card or passport. For example, in Germany, only people older than 16 years old are required by law to have such an identity document.
- *Tracking and Profiling.* The procedures of age verification are potentially privacy invasive. Depending on the choice of architecture (see online assurance in 4.3), the IdP or AAS might learn which services a user is accessing when.

*Risk-Benefit-Analysis (document-based).* For document-based verification the following additional risks exist.

- *Information Leakage Towards Verifier.* In almost all cases the verifying party directly learns the content of a user's official identity document, as well as, often, a photo or video of the person. The information contained in official documents go well beyond what is needed for age verification. Contrary to the relationship-based approach, the verifying party would typically not already have access to these data. This exposes users to security and privacy-related risks, such as data leaks[8] or profiling.

*Risk-Benefit-Analysis (relation-based).* For relationship-based verification, the following additional risks exist.

- *Lack of Data Minimisation.* Depending on the specific method of sharing age verification data, the identity provider might transfer more information than is strictly necessary to the service provider. For example, maybe the data includes a user's name or username, or the full data of birth instead of an age range.

### 3.2.3 Age Verification: Acceptance

The empirical studies assessed for this article focussed on representative surveys, most of which had been conducted either in the context of planned legislation in Australia and UK in 2021-2024 or in the wider context of regulatory debates regarding age restrictions of adult content services. In these surveys, age verification via official sources of identification (e.g. government ID, credit card details) is seen as a straightforward and reliable way to prove age. The public clearly recognises the potential effectiveness of ID-based verification in theory: showing an official ID is analogous to how age is confirmed for alcohol or tobacco purchases offline, and it is intuitively understood as a reliable gatekeeper.

However, public trust in *how* these verification processes are implemented is much more tenuous. Support for the general concept of age verification often evaporates when people are asked to actually provide their own personal data online, as many users are uncomfortable sharing sensitive identity information with private websites or apps. A U.S. survey found that two-thirds of Americans are not comfortable uploading a government ID to social media companies to verify age (Barkley, 2023); even more (70%) are unwilling to provide their child's ID for age verification purposes. This discomfort is echoed in other countries: in an UK poll, large majorities said they support age checks in principle, but only 37% would actually be willing to show proof of age for social media, and a mere 19% for dating apps (Ipsos, 2025). Thus, many adults appear to endorse age verification rules for others, yet balk at the intrusiveness when it applies to themselves.

The public's hesitation seems to stem from a variety of concerns that are being mentioned in surveys. Here, privacy and data security are top of mind: people fear that uploading IDs or credit card data could expose them to identity theft or data breaches. In Australia's trials, experts stressed that any centralized database of IDs could become a high-value target for hackers; they also worry whether companies will properly delete personal documents after verification (Scimex, 2025). Indeed, trust in online platforms to safely handle personal data is generally low – less than 5% of Australian adults fully trust platforms to securely store their information, while almost half have "no trust at all" (SRC, 2024). Many respondents indicate they would trust government agencies or other official third parties more than private companies to manage age verification data (eSafety Commissioner, 2021). For instance, Australia's eSafety survey found people felt the government was "best placed" to process age verification info securely and ensure the system works as intended (eSafety Commissioner, 2021).

---

[8] See for example https://cybernews.com/security/global-data-leak-exposes-billion-records/ which reports a large data leak involving identity data.

*Perceived effectiveness* of ID-based age verification reveals a mixed picture. On one hand, people acknowledge that checking an official ID is highly reliable in confirming age (SRC, 2024). In surveys, the primary perceived benefit of these measures – cited by around 62% of Australian adults – is the added safeguard for children, ensuring only adults gain access to adult content (eSafety Commissioner, 2021). On the other hand, sizable segments of the public doubt how effective these checks will be in practice: About 24% of a survey's respondents lacked confidence in the practical effectiveness of porn site age verification, raising issues such as the ease of bypassing or tricking the system (mentioned by 28%) and the possibility of youths using parents' or borrowed credentials. There is widespread scepticism that tech-savvy teens will find workarounds (e.g., using VPNs, fake IDs, or stolen logins), thus rendering such age gates obsolete. Indeed, reports from early implementations indicate some minors have already hacked the system (Muntinga, 2025). These stories might fuel public doubt that age verification will ever be foolproof. Moreover, people worry about exclusion and fairness: any system that requires formal ID could wrongfully block those who lack identification documents.

Public opinion also varies by region in trusting ID-based solutions. Countries with established digital identity systems or stronger privacy laws see relatively higher acceptance. A global survey in 2023 found 68% of consumers worldwide were open to using a digital ID for online verification, but this drops to 62% in the U.S. and 61% in the UK (Jumio, 2023). Experts suggest Americans in particular are less inclined to trust technology companies with personal data due to the lack of a broad national privacy law (Teale, 2024). By contrast, in countries like Singapore or Mexico, public openness to digital identity verification is higher (around 78% and 71% respectively (Jumio, 2023), perhaps reflecting different cultural norms or government initiatives in digital ID.

Despite these differences, a commonly mentioned thread is that any age verification regime must convincingly safeguard user data. If people feel that verifying their age online is comparable to surrendering their entire identity or exposing themselves to surveillance, resistance remains high. Conversely, approaches like cryptographic age tokens which prove one is adult without revealing other personal details are viewed much more favorably; in one Australian study, such "confirmation tokens" were rated the most comfortable age assurance method across all age groups (SRC, 2024).

## 3.3 Age Estimation

### 3.3.1 Age Estimation: Technological Foundations

Age estimation uses biometric data from users to predict their age using machine learning techniques. The most common form is facial age estimation, where a user's face image is analyzed to determine an approximate age, i.e. if the user *looks* like a minor or an adult. Other options are voice-based age estimation, which analyze speech patterns, or gesture analysis. Each option requires the collection of relevant features in order to do estimation. For example, users might be asked to take a photo or video of their face with their device's camera, a machine learning approach then uses this data to estimate the user's age.

### 3.3.2 Age Estimation: Analysis

*Effectiveness & Assurance Level.* We rate the assurance level of age estimation as low-medium, as early deployments of facial age estimation have shown accuracy limitations: In the UK, regulators noted that "age assurance technologies which scan your face and estimate your age don't work very well on children" in certain cases (McConvey, 2024). The *accuracy* tends to be poorest for teens and young adults, precisely the group of interest for age checks.

Independent tests by the U.S. NIST found that even the best facial age algorithms in 2024 had a mean absolute error of around 3 to 5 years for teen subjects (NIST, 2024). This margin of error is significant: it means a typical 17-year-old might be misclassified as 20+ or as 13-14 in worst case scenarios, leading to unwanted false negatives or false positives. Indeed, in NIST's (2024) study, fewer than 35% of 13-year-olds were correctly estimated within one year of their true age by any algorithm tested.

*Risk-Benefit-Analysis.* On the positive side, age estimation is more inclusive in comparison to age verification and provides convenience:

- It does not rely on any existing identity documents, and thus is more inclusive for younger people without such documents.
- Moreover, age estimation solutions offer convenience: instead of having to upload documents, users might simply look into a camera or speak into the microphone.

On the negative side, there might be some risks regarding bias and privacy:

- *Risk of bias.* While we noted that age estimation may be more inclusive by not excluding people without existing IDs, it can also be less inclusive as data-based systems tend to be biased. Biases and differences regarding accuracy could also apply along other dimensions, such as gender, facial deformities, or skin colour.
- *Risk for Privacy.* The privacy risks of age estimation are somewhat controversial: While some proponents consider this approach to be privacy-friendly (AAT, 2025, p. 76, A.25.4), we reach a different verdict: biometric data are highly sensitive and submitting them for age assurance purposes thus should not be considered privacy-friendly in general.[9] Photos or recordings submitted for the purpose of age estimation are typically not yet known to the verifier. Worse, requiring age assurance would indeed incentivize such biometric data analysis or may even make it obligatory.

### 3.3.3 Age Estimation: Acceptance

Industry proponents market AI-based age estimation as a privacy-friendly alternative, since it can verify age without storing personal IDs or requiring real names (TechUK, 2024). Public awareness of these tools, however, remains limited. In 2021, only 18% of Australian adults surveyed were even aware of the concept of "age estimation" technology before being prompted (eSafety Commissioner, 2021). Once explained, the idea often elicits cautious interest coupled with concerns about accuracy and privacy. Public trust in AI age estimation is lukewarm at best, and generally lower than trust in traditional ID checks (CETaS, 2025; Lai, 2025; Forland et al., 2024). For example, a large 2024 Australian study found that only about 38% of adults felt comfortable with sharing a face scan or voice print to prove their age online (SRC, 2024); in contrast, two-thirds were comfortable showing an official ID card for age verification. This indicates that biometric age checks are still viewed warily by a majority. Qualitative feedback reveals why: experts claim that many people are unnerved by facial recognition technology and unsure if their images will truly be deleted or kept secure (Scimex, 2025). Unlike flashing an ID to a clerk in person, submitting a selfie or turning on the webcam for AI analysis means relinquishing an image of oneself to an algorithm and whoever operates it. This still unfamiliar process seems to invite skepticism.

---

[9] We do recognize that procedural measures and external audits can limit the impact of privacy leaks. However, these audits and the underlying software are usually not verifiable.

Another major public concern is whether AI age estimation actually works *reliably*, especially for edge cases. People intuitively understand that guessing age from appearance is prone to error – after all, humans often misjudge others' ages, and they suspect machines can err too. Assessment of the effectiveness of facial age estimation (see 3.3.2 above) might have reinforced public scepticism. There are fears that underage users will slip through if the AI underestimates their age by a few years, or conversely that older users will be wrongly blocked if the AI estimates their age as too low (Scimex, 2025; Wagener, 2025). Both types of errors may undermine confidence in the technology. Media reports of teens fooling facial age checks (for instance, by using photos of older people or even AI-generated deepfake faces) further dent the method's perceived effectiveness (McIlroy-Young, 2015). Survey data reflect these doubts, as Australia's trial has shown that "biometric information" was among the least trusted approaches, with young participants especially uncomfortable at the idea of an AI scanning their face; respondents questioned the legitimacy of such scans, voicing that it did not "seem trustworthy" compared to more familiar ID checks (SRC, 2024).

Another key public concern is privacy and ethics: Facial age estimation requires capturing an image or a video of the user's face, not profoundly different from face recognition, a technology that has prompted privacy backlash around the world. Experts worry that a database of biometrical face scans could be misused, hacked, or repurposed for surveillance beyond age checking (Scimex, 2025). Although providers like Yoti, a leading vendor of AI age estimation services, assert that they instantly delete images and only output an age prediction, users often remain wary, lacking a way to verify these claims. The sensitivity of biometric data amplifies fears: unlike a password, one's face or voice cannot easily be changed if compromised. In surveys, participants have voiced concern that a breach of a facial age estimation system would be especially damaging, potentially exposing "more diverse and more sensitive types of personal information" than even an ID leak (SRC, 2024).

Moreover, bias and fairness issues are being mentioned: AI algorithms have been documented to perform unevenly across different demographic groups. For instance, some facial age models are less accurate for females and people of colour (Kotwal & Marcel, 2025; Puc et al., 2020), which could result in systematic disparities – e.g. some groups being asked for additional verification more often due to higher error rates. Such bias would erode public trust further, as users expect fair and equal treatment. The industry is working to improve accuracy and reduce bias, e.g. tuning algorithms on diverse data (Trotman, 2024), and results are improving. For example, Yoti reports their latest system is highly accurate for ages 13–17, with 99% true positive rate in lab conditions (Yoti, 2024). But many regulators and civil society groups remain cautious. A 2022 audit by France's data protection authority (CNIL) concluded that no current age estimation solution fully met the required standards for reliability and privacy simultaneously (CNIL, 2022).

Public acceptance of age-estimating AI remains tentative. When forced to choose, some users do prefer a quick face scan over uploading an ID, especially if they value anonymity. But this is often seen as the lesser of two evils. Notably, the Australian study found people were more comfortable with facial estimation than giving a credit card or ID document, yet still less comfortable than with non-biometric methods (SRC, 2024).

## 3.4 Age Inference

### 3.4.1 Age Inference: Technological Foundations

Age inference technologies attempt to deduce a user's age *indirectly* by analysing their digital behaviour, metadata, or other signals. Rather than asking the user to present evidence, these methods operate in the background – for example, monitoring one's browsing history, social media interactions including uploaded texts, pictures or videos, search queries, or game play patterns to judge if the user *behaves* like a minor or an adult. Some platforms already employ AI-driven inference to flag accounts that may belong to underage users (for instance, scanning posts or friend networks for signs of a child's presence).

It should be noted that age inference technologies encompass a wide range of different solutions. Using this umbrella term therefore may hide important distinctions. One crucial distinction for instance refers to the *type and origin of data* used for the inference. Some depictions of AI seem to imply that platforms doing age inference use only the data they obtain about their users anyway. While the massiveness and sensitivity of this data alone should not be underestimated (after all, how much and how granular data would be needed to reliably distinguish younger persons), other accounts take a considerably wider view. The Australian technology trial, for instance, widens this scope and also considers the combination of platform data with data from external sources, such as electoral enrollment, school year, or email history (AAT, 2025, p. 91, A.31.2).

### 3.4.2 Age Inference: Analysis

*Effectiveness & Assurance Level.* No clear data exists on the effectiveness of age inference. Depending on the strength of the signals, it is often unlikely that behavioural data enables easy distinction between, for example, a 15 and a 16 year old teenager (AAT, 2025). We therefore rate the assurance level as low. Indeed, experts have indicated, age estimation might be more useful as a signal that a more precise age check is needed.

*Risk-Benefit-Analysis.* Age Estimation is not applicable in all contexts. It requires significant data about users to even be able to perform a reasonable assessment of a user's age. This data is likely present on social media platforms and LLM platforms, but it might not be present on regular websites nor porn websites.

On the positive side, age inference does not need checking and interaction with users:

- *Frictionless.* Age inference does not, typically, require users to provide additional information, platforms simply use the information that is already available to them.
- *Likely difficult to circumvent.* Age inference is bound to an entire account's history; something that cannot easily be changed by a friendly relative showing up (see General limits of Technology in Section 5.1).
- *Allows for continuous checking.* Age inference can be applied in an "always-on" manner, and can therefore detect if a (younger) person is using an account created by an older user. Subsequently, it could trigger age estimation or age verification in situations where the system has detected unexpected behaviour.

On the negative side, age inference is very invasive to the user's privacy and has some limitations:

- *Normalizes surveillance.* Age inference risks normalizing the idea that it is acceptable to surveil users by collecting large amounts of data about them.

- *Increased data collection.* Age inference likely works better when more data is gathered about users of a platform. As such, age inference approaches run the strong risk of increasing data collection and storage about users, and to serve as an excuse why data minimization and a strong limitation of processing purposes are not possible.
- *Push to create accounts.* When age inference is one of the main age assurance technologies, social media platforms and other service providers might push users to create accounts to facilitate inference, potentially leading to increased tracking and profiling.
- *Does not work from the beginning.* Age inference requires the collection of sufficient user data before it can be effectively applied. This may incentivize the creation of numerous short-lived accounts that are used only briefly, before age inference mechanisms begin to function reliably for a given account.

### 3.4.3 Age Inference: Acceptance

From a usability as well as from a technical perspective, age inference is attractive because it can be frictionless for users. As no prompts or uploads are required, the technology is practically "invisible". However, in surveys, this approach is by far the most controversial: When explained to users, it often triggers immediate discomfort about "Big Brother" style surveillance. In an Australian survey, "activity profiling" was consistently rated the least acceptable age assurance method among both adults and teens (SRC, 2024); only about 16% of adults felt comfortable with the idea of technology that tracks their online activity to guess their age – essentially an overwhelming 84% were uneasy or opposed. This was on par with or even worse than the level of discomfort expressed about sharing credit card details (also around 16% who are comfortable).

Qualitative feedback reveals that people see age inference as inherently intrusive. Participants in focus groups voiced that such continuous tracking "constitutes a breach of personal privacy" by its very nature (SRC, 2024). The notion that a company would be "watching what we do online over long periods to guess our age" has been described as a "significant shift toward surveillance" that most found unacceptable (Scimex, 2025). Even if the intention (protecting kids) is good, the method of constant monitoring clashes with users' expectations of privacy and data minimisation. This is somewhat remarkable, since platform providers usually reserve their right to comprehensively track and profile all their users for optimal ad-related segmentation anyway.

Public trust in the accuracy of age inference is also extremely low. Unlike presenting an ID or scanning a face which at least directly measures an attribute of the user, behavioural inference seems indirect and error-prone to many. People have diverse online behaviours, and it is not clear to the public how reliably one's age can be divined from, say, the videos they watch or the slang they use in chats. Focus group respondents in one study highlighted a "lack of robustness and accuracy" in any method that tries to assess age from activity patterns (SRC, 2024). They easily imagine scenarios where an adult might resemble a teen in behaviour or vice versa, leading to mistakes. False classifications could have real consequences – e.g. an adult locked out of their account or a minor not flagged at all – which is undermining confidence in the system's fairness.

The opaque nature of these algorithms further erodes trust: with ID checks or face scans, a user at least knows they are being evaluated at a given moment. But with inference, the criteria are hidden and decisions might be made without the user's awareness or consent. This lack of transparency breeds suspicion that the algorithms could be wrong and users would have little recourse to correct mistakes. Moreover, ethical concerns are looming: Behavioural age inference often relies on aggregating personal data, essentially building a digital profile of one's activities. Users worry this could open the door to even broader profiling and data misuse. There is a slippery-slope fear: a system designed to infer age might incidentally reveal or exploit other personal attributes (interests, habits, even sexual or psychological traits) for commercial gain. Already, targeted advertising uses behavioural profiles. Extending that to age verification raises red flags in many with regard to the so-called "function creep". In Europe, data protection authorities have cautioned that any reuse of collected data for age inference must be done with great care and clear legal basis, given the privacy implications (CNIL, 2022); the legal overview above has shown that the DSA is limiting the use of additional data processed for age verification. Many in the public likely would not consent to such extensive monitoring solely for age checks, especially when it happens behind the scenes. Notably, younger users (who are the supposed beneficiaries of these protections) also find the idea objectionable: In an Australian study, teens expressed that covert monitoring of their online life to enforce age rules feels invasive and undermines trust in platforms (SRC, 2024).

Due to these factors, age inference seems to be the least publicly accepted of the age assurance approaches. It tends to be viewed as a "last resort" or a supplementary measure rather than a primary solution. Some experts have suggested that inference could be used in a limited way – for example, to quietly prompt a user to re-verify their age if the system detects strong signals of mismatch (like childlike behaviour on an adult account). Framed this way, it might be slightly more acceptable as an internal safety net. But as a standalone method to enforce age restrictions, it currently faces steep public opposition.

In policymaker consultations, privacy advocates have argued that broad adoption of behavioural age inference would normalize surveillance and erode online anonymity for all users, with only marginal gains for child safety (SRC, 2024; Scimex, 2025). These arguments resonate with a public increasingly concerned about digital privacy. Indeed, when given alternatives, survey participants overwhelmingly prefer age checks that do not require tracking their every click. Australian consumer research found that any method involving "tracking online activity" ranked at the bottom for user comfort (only 16% comfortable), whereas methods like single-use age tokens or one-time ID checks were far more accepted (SRC, 2024). People value their privacy and autonomy online, and age inference is perceived as violating those values; unless implemented with extraordinary transparency, strict data limits, and user control, it will likely remain an unpopular approach.

## 3.5 Parental Control and Consent

### 3.5.1 Parental Control and Consent: Technological Foundations

When using *parental control* technology, parents provision a child's device, and specify what the user of the device can access and what not. From our understanding, this can be seen as an AAT since the parents either configure the system according to the age of the child, or they put the actual age of the child as information in parental control software (as a kind of parental age vouching). Apple provides iOS Screen Time to limit what children can do on their devices, while Google provides Google Family Link to control a child's device. Optionally, the devices could be configured to request *parental consent*. Parents would be asked to approve access when the child tries to access restricted or unknown content.

### 3.5.2 Parental Control and Consent: Analysis

*Effectiveness & Assurance Level.* Assuming that parents apply parental control correctly on their children's devices, this approach is effective in preventing children from accessing age restricted resources and apps. Moreover, circumvention is also made more difficult, as parental controls on mobile operating systems can also limit the use of VPN apps, a common tool to circumvent age restrictions.

However, this high level of assurance is only obtained when parents actively and correctly set up parental controls.

The available evidence on whether platform-level parental controls reduce problematic social media use is mixed and, with regard to the few empirical studies, discouraging: An internal Meta study ("Project Myst"), reported in February 2026, surveyed 1,000 teens and their parents and concluded that in-app supervision tools such as time limits and access restrictions had little measurable impact on teens' compulsive social media use. The study found no clear association between parental supervision and adolescents' capacity for self-regulation, while stressful life events and individual vulnerability factors were more strongly predictive of overuse (Perez, 2026).

Broader longitudinal research on restrictive mediation affirms this scepticism. Imposing strict rules regarding the amount, timing, or location of social media use does not reliably prevent problematic engagement and can, in certain contexts, be associated with more negative attitudes or increased engagement with restricted content outside parental view. A mixed-methods study of early adolescents found that restrictive monitoring was positively associated with parents' perceptions of problematic internet use, whereas discussion-based monitoring was embedded in more positive family dynamics (Hernandez et al., 2024; Sevilla-Fernández et al., 2025). Research on TikTok specifically suggests that active, discussion-based parental mediation can moderate the link between intense platform engagement and negative outcomes more effectively than technical restrictions alone (Qin, Musetti & Omar, 2023).

A consistent finding across the literature is that platform-level parental controls function as one tool within a broader, multimodal mediation repertoire rather than as standalone solutions. Drawing on parental mediation theory, studies differentiate active mediation (conversation, co-use, guidance), restrictive mediation (rules, filters, monitoring), and enabling mediation (supporting digital skills and exploration). Technical controls on social media constitute a subset of restrictive mediation. Recent reviews emphasise that parents dynamically shift between strategies and that the uptake of platform-specific supervision tools depends substantially on family communication patterns, adolescent age and autonomy, and the perceived quality of the parent-child relationship (Rodríguez-de-Dios et al., 2018; Daneels & Vanwynsberghe, 2017; Sevilla-Fernández et al., 2025; Livingstone et al., 2015).

*Risk-Benefit-Analysis.* Parental control & consent has the following benefits:

- *Parents stay in control.* Parents can exercise their parental rights, and can make individual decisions that they deem are appropriate for their children.
- *Resistant to circumvention.* While age verification and age estimation are susceptible to circumvention by adults (see generic risks in 5.1), parental control locks down a child's device making circumvention harder (e.g., VPNs apps cannot be installed anymore).

On the other side, the following risks exist:

- *Only effective if enabled by parents.* Parental control and consent is ineffective when not actively set up by parents or disabled, e.g. upon pressure by children themselves.
- *Inequalities and biases.* Parental Control and Consent (PCC) may reinforce societal inequalities and biases: First, children whose parents do either not care or do not bear up against pressure remain unprotected. Second, children with special needs or interests, e.g. LGBTQ+ children and youth who seek support which they don't want their parents to be aware of, may be harmed.

### 3.5.3 Parental Control and Consent: Acceptance

Surveys consistently reveal a pronounced gap between high levels of parental concern about social media risks and comparatively low uptake of available control features. Ipsos data show that while approximately 70% of parents are aware of parental controls on video-streaming services, awareness drops to around 43% for video- and image-sharing platforms and gaming apps (Ipsos, 2023). Actual use lags further behind: only about one in four parents report using controls on microblogging apps, and roughly one in five on chat- or thread-based services (Ipsos, 2023). A Family Online Safety Institute report found that fewer than half of parents (approximately 47%) fully utilise the controls available on their children's smartphones, despite growing feature sets on Instagram and TikTok (FOSI 2025). Notably, 60% of non-users state that they refrain from using controls because they trust their children to make good decisions online (Ipsos, 2023), indicating that low adoption reflects deliberate parental choice as much as unawareness or technical barriers.

The research identifies several recurring barriers that limit acceptance of social media parental controls. First, trust and relational concerns figure prominently: parents fear that activating supervision features signals distrust and may damage the parent-child relationship, particularly with older adolescents (Ipsos, 2023). Second, usability and literacy barriers are significant. FOSI data and qualitative studies report that parents find controls difficult to locate, understand, or configure, with some parents ironically asking their children for help setting up the tools meant to supervise them. Third, effectiveness doubts and circumvention play a major role. Young users frequently learn to bypass restrictions, leading parents to question whether the configuration effort is worthwhile. Qualitative work documents that parental strategies are in constant flux as families experiment with different approaches amid uncertainty about what actually works (Hernandez et al., 2023). Also, privacy concerns are relevant barriers as well: survey data indicate that 44% of parents worry that using controls invades their child's privacy (MacBook Journal, 2024).

## 3.6 Summary and Preliminary Conclusions

Table 3.1 provides a summary of the age assurance approaches, their respective assurance levels, side effects, acceptance, and their applicability. We refer to the detailed discussion above for the details.

|  | Assurance Level | Side Effects | Acceptance | Applicability |
|---|---|---|---|---|
| **Age Verification** | High | High | High | Broad |
| **Age Estimation** | Low-Medium | Medium-High | Low | Broad |
| **Age Inference** | Low-Medium | Medium-High | Low | Limited to large social platforms with accounts |
| **Parental Control** | High (if enabled) | Low | Medium | Limited to specific child devices |

*Table 3.1: Summary of assurance approaches*

# 4 Age Assurance Architectures

In addition to the four different approaches to age assurance presented in Section 3, there are also three different architectural options to transfer the results of an age assurance method for a user to the service provider (SP), such as a social media platform or a porn website. These architectures involve different actors outlined below. Moreover, these architectural choices influence effectiveness, ease of deployability as well as security and privacy risks.

## 4.1 Three Different Age Assurance Architectures

We distinguish three different architectures for transferring the results of an age assurance method to the service provider (SP), such as a social media platform or a porn website (see also Figure 4.1).

1. Online Assurance. In online age assurance mode, the user performs the age assurance process right before accessing restricted resources at the SP. The user interacts with the age assurance service (AAS), the Identity Provider (IdP), or directly with the Service Provider (SP). In any case, the result of the age assurance process is sent to the SP.
2. Offline Assurance – Device-Based. In offline assurance, the assurance process only happens once (or infrequently). In device-based offline assurance, the user's device performs a one-time assessment of the user's age, and stores the result on the device. Whenever a user wants to access a restricted resource at an SP, the device sends along an "age claim" (a statement of the user's age) to enable access.
3. Offline Assurance – Credential-Based. In credential-based offline assurance, the user interacts with an age assurance service (AAS) or identity provider (IdP) only once to obtain a credential that attests to the user's age. Once this is obtained, the device sends, upon request, a cryptographic "age proof" that attests to the age of the owner of this device, whenever the user attempts to access a restricted resource at a service provider (SP).

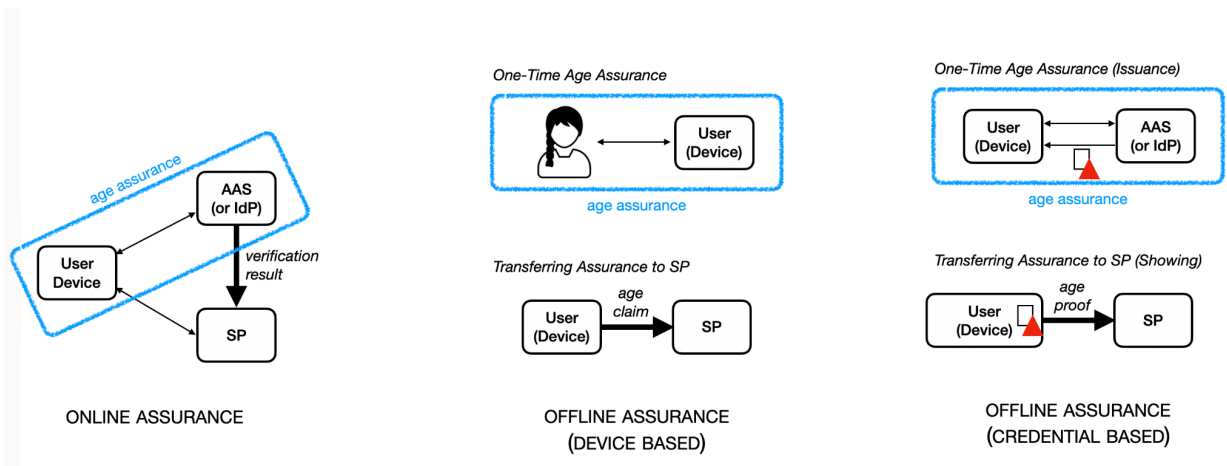

*Figure 4.1: Different Age Assurance Architectures*

## 4.2 Actors in Age Assurance Architectures

We briefly summarise the roles of different parties that play a role in the age assurance architectures we discuss.

**User.** A user wishing to access age restricted content. They might do so from a desktop computer or a smartphone.

**Service Provider (SP).** A service provider offering age restricted content, i.e. a social network, a website, or a smart phone application. A service provider might operate a website, a platform for a smartphone app, or both.

We furthermore consider the following parties that directly or indirectly provide information about a user.

**Identity Provider (IdP).** An identity provider has an existing relationship with a user, such as a government authority, a bank, or a (health)-insurance company. For the purpose of this document, we assume that (1) identity providers have the means to establish a user's identity with high confidence (e.g., through existing authentication methods) and (2) they know the user's date of birth or age with sufficient accuracy to enable establishing a user's age.

Finally, we foresee the use of a third party *age assurance service* that plays a role in the age assurance process.

**Age Assurance Service (AAS).** These commercial entities provide age assurance as a service. For example, instead of a service provider (say a porn website) directly verifying a user's identity document, they might rely on a third party to do this for them. We call these third parties *age assurance services.*

We expect that age assurance services will play a major role in any deployed age assurance system. Most age assurance approaches we have discussed are highly technical, and require expertise to execute well. We therefore expect that commercial entities will fill this gap and provide age assurance as a service.

*Compatibility between assurance approaches and architectures.* These architectures are mostly compatible with each of the different assurance approaches we discussed before (see Table 3.1), but there are two notable exceptions. As we explain below, age inference requires more information, making it less suitable for some combinations of approaches and architectures, whereas parental control effectively implies a device-based approach.

|  | Age Verification | Age Estimation | Age Inference | Parental Control |
|---|---|---|---|---|
| Online Assurance | Yes/possible | Yes/possible | Yes/possibly (if AAS knows user) | Not possible |
| Offline Assurance (Device-Based) | Yes/possible | Yes/possible | Possibly only if OS uses on-device behavior | Yes/Possible (but not needed) |
| Offline Assurance (Credential-Based) | Yes/possible | Yes/possible | No/impossible (unless platform acts as IdP) | Not applicable |

**Table 4.2: Possibility of combining approaches (rows) and architectures (columns)**

## 4.3 Online Assurance

### 4.3.1 Online Assurance: Technological Foundations

In the online assurance architecture, the age assurance check is performed online, i.e. in real time right before the user accesses an age restricted resource. The typical workflow in online assurance is as follows:

1. When requesting an age restricted resource at a service provider, the service provider (SP) requests an age verification from a third party: an age assurance service (AAS) or identity provider (IdP).
2. The third party (AAS or IdP) then interacts with the user and applies the age assurance approach to obtain evidence of the user's age.
3. The third party sends the result back to the service provider, and the user gains access to the requested resource.

We expect that the use of a third party (AAS or IdP) for age assurance purposes will be by far the most common scenario. First, this fits modern service architectures where separate parties provide specific services that are only lightly coupled via application programming interfaces (APIs). Second, implementing age assurance approaches is in most cases difficult, reinforcing the expectation that third parties will emerge to provide age assurance as a service. In fact, a whole industry is currently emerging; a market report from 2024 expects the turnover of the global age assurance industry to be around 10 billion US-$ by 2029 (Liminal 2024).

This information flow is very similar to Single-Sign On (SSO) solutions, where the user is redirected to their IdP, logs in there, and then is redirected to the service provider. Popular versions of SSO-based solutions include Facebook Login, Log-in with Apple, Google Sign-In.

*Compatibility.* The online assurance approach is directly compatible with age verification and age estimation. Online assurance is also compatible with age inference, but only when the third party has sufficient data about the user to infer their age from this data. This is for example the case when a social network acts as an IdP or AAS in the above process.

Finally, online verification is not compatible with parental control and consent. First, the third party has no ability to look up the parents control settings. Second, it is very difficult to implement asking for parental consent. This would entail: (1) identifying the child; (2) looking up the legal parents and/or guardians; (3) verifying the identity of the parent/guardian that is present; and (4) finally asking for consent. Not only is this an invasive process, especially the parent/guardian-child relation is very difficult to verify in digital environments; in many countries, it is rarely captured in official documents.

### 4.3.2 Online Assurance: Analysis

Pros:

- When used in an account-less setting, i.e. where the service provider (SP) does not store the verification result, online assurance is harder to circumvent. Since the SP does not store the result, assurance needs to be executed frequently. Thus "helpful" adults would need to be present every time a restricted resource is accessed.

Cons:

- Privacy leakage: The third party (AAS or IdP) typically learns which service provider the user is accessing. This leakage of information about users' behaviour is very difficult – if not impossible – to avoid in online assurance architectures. In fact, it is the very fact that assurance and access are contemporary that makes it impossible to fully remove this leakage.

## 4.4 Offline Assurance

In an offline age assurance architecture the time of age assurance is decoupled from the moment a user accesses an age restricted resource. In between, the result is stored on a user's device. We distinguish two methods of storage, one where the device is trusted (device-based) and one where the device does not need to be trusted, but instead provides cryptographic proof that age assurance has happened (credential-based).

Pro:

- The big advantage of offline assurance is that the inherent leakage of a user's online activities towards the AAS or IdP are no longer an immediate problem. In fact, the whole point of offline assurance is that the IdP or AAS is *not* involved in the actual accessing of restricted resources.

Con:

- Offline assurance is easier to circumvent durably. Offline assurance stores results for some time. As a result, it is more easy to durably circumvent age limits. All a child needs to do is to get one older person to perform the age assurance method on their behalf. Once successful, the result is stored, and can be used for a longer time.

We emphasize that this downside is not the result of any technical choice in storage method. Stronger cryptography, or more reliable assurance methods are of very little help, except when these assurance methods can *detect* that the person performing the age assurance is *not* the normal user of the device (e.g., by applying continuous age estimation or inference).

## 4.5 Offline Assurance – Device-Based

### 4.5.1 Offline Assurance – Device-Based: Technological Foundations

In device-based offline assurance, the user's device performs the age assurance, either directly or by relying on a third party assurance service. This is in line with recent news that Apple is experimenting with performing age assurance directly at the iOS level (Apple, 2026). And a recent bill in California requires operating systems to collect age data from the use upon registration.[10]

---

[10] https://leginfo.legislature.ca.gov/faces/billTextClient.xhtml?bill_id=202520260AB1043

A device-based offline assurance architecture functions as follows.

1. *One-time age assurance.* Once (or very rarely), the user's device directly executes an age assessment with its user. Ideally this assessment is performed directly on the device.[11] The device's operating system, a wallet, or an app on the device permanently stores the result of this assessment.
2. *Transferring assurance.* Whenever a user attempts to access an age restricted resource, the device simply informs the service provider of the result of the assessment and the service provider trusts whatever the device says.

*Compatibility.* Device-based offline assurance is directly compatible with age verification and age estimation. Both can be executed directly on the device. In theory, age inference is also possible, as the device has access to a huge trove of user behaviour data. However, this does require a much more complex implementation on device, and is likely to result in push-back from users that this is too invasive.

Finally, device-based offline assurance is compatible with parental control, but no integration is needed: parental control methods can directly restrict access without informing the service provider of the user's age.

### 4.5.2 Offline Assurance – Device-Based: Analysis

Pros:

- When age assurance happens on the user's device, the privacy risks of age verification, age estimation, and age inference disappear. The device likely already has access to all these data; so no new risks appear.
- This approach is very easy to implement for service providers, they only need to honor the device's signal about the user's age.

Cons:

- The enforcement is more diluted. While service providers of restricted resources can easily check *that* they received an age assessment result from the user's device; they cannot determine whether this assessment is valid. (The device might respond with invalid age information.)

---

[11] Devices can execute any age assurance approach. We assume in the following that this approach is executed on the device itself (e.g., for age verification, the device reads an identity document and locally checks the holder's age; for age estimation, the device takes a photo and locally estimates the age). However, devices could also rely on external age assurance services to perform age assurance, and then store the result. Since the latter has weaker privacy properties than local age assurance, we do not further consider it here.

- The very strict requirements may have detrimental effects on open source providers, thereby endangering open source platforms, and alternative mobile operating systems. See also the risk of collateral damage in 5.2.4.

## 4.6 Offline Assurance – Credential-Based

### 4.6.1 Offline Assurance – Credential-Based: Technological Foundations

In credential-based offline assurance, the moment of assurance and the moment of accessing restricted resources is again decoupled. This time, however, the device stores a cryptographic proof that age assessment happened. In other words: the device can no longer convey incorrect information.

Credential-based architectures follow the following high-level process:

1. *One-Time Age Assurance (Issuance).* The user's device interacts with an age assurance service (AAS) or identity provider (IdP).
   First the AAS/IdP establishes the user's age. For example, when using age estimation, the AAS might request a selfie and use this to approximate the user's age. An IdP on the other hand, might ask a user to log in, so that the IdP can look-up the user's age. Once the age has been established, the AAS/IdP issues an age credential to the user's device. The device stores this credential in a wallet.
2. *Transferring assurance.* Whenever a user attempts to access an age restricted resource, the user uses their credential wallet to construct a cryptographic proof of the user's age; and sends this to the service provider.

There are different ways that credentials can be instantiated. One example is the new EU Digital Identity Framework, following the revised eIDAS 2.0 regulation, which proposes an EU wide digital wallet that enables issuing and verification of credentials across Europe.

Digital credentials are in many ways the digital equivalent to identity documents. As with identity documents, credentials are issued by a trusted party, the identity provider (IdP) or age assurance service (AAS). Like an identity document, digital credentials can contain many fields of information, traditionally called attributes, which together are attested to or signed by the IdP.

Physical identity documents can be shown to anyone to verify some information, for example that the holder is over 18 years old. The verifier can rely on the information contained in the document as long as the verifier trusts the issuer of the document. When shown to people rather than being read electronically, identity documents also provide a reasonable measure of privacy. While a lot of information is available on the document, it is difficult for human verifiers to remember this information.

Digital credentials should replicate these privacy-features for digital verifiers while still guaranteeing authenticity of the information contained within the credential. Cryptographers and privacy researchers have identified 3 properties that are essential for digital credentials to provide privacy:

- *Selective Disclosure.* Even when a credential contains many pieces of information, it should be possible for a device to reveal only the relevant ones. For age assurance, this means that when a credential also contains a name, or date of birth, it should be possible to only show to the service provider *that the user is old enough*, without revealing any other information about the name or date of birth.
*Prevented harm:* Selective disclosure helps prevent profiling and tracking of users. Selective disclosure hides all user-specific information except for age. Thus it is not possible to use this information to recognize (and thus track users) or help build a profile about this user.
- *Issuer Unlinkability.* It should not be possible for the issuer (IdP or AAS) to recognize *where* a credential is being used. This directly prevents the downside that were identified for the online architecture above.
*Prevented harm:* Identity providers (and age assurance services) cannot build a record of a user's behaviour online. Concretely, this for example prevents a bank (acting as the IdP) from learning that one of their customers is visiting a gambling website.
- *Verifier Unlinkability.* Verifiers or service providers should not be able to recognize users based on the cryptographic proofs it receives.
*Prevented harm:* Verifier unlinkability ensures that service providers cannot profile a user's behaviour as a result of the digital credentials.

The EUDIF framework requires all three properties to be satisfied. We highlight three approaches that are often mentioned in the context of age verification:

1. *Attribute-based Credentials.* These are full-fledged digital credentials that satisfy all the properties stated above, and form ideal candidates for implementing the EU Digital Identity Framework. Different cryptographic solutions have been proposed over the years, starting with Idemix and U-Prove in the early 2000s (Camenisch & Lysyanskaya, 2001; Brands, 2000). These and later approaches have been well-vetted by the cryptographic community. Underlying all of these approaches are so-called *zero-knowledge proofs*: they enable the user's wallet to create a proof that the user has a valid credential (the "proof" part), while hiding all information that is not strictly necessary (the "zero-knowledge" part).
2. *Batch issuance.* To simplify and speed up the deployment, recent proposals have instead looked at batch issuing one-time use tokens. These are much simpler than attribute-based signatures. One can think of these as a voucher that says "the holder of this voucher with serial number XYZ is over 18 years old". As a result, these effectively achieve selective disclosure, and provide verifier unlinkability provided that each voucher is used only once. This approach, however, does *not provide issuer unlinkability*, and thus has less privacy protections.
3. *Skipping the Issuance Step.* A third approach directly uses the digital information stored in modern passports and ID cards to create a cryptographic proof, thus skipping the one-time verification step. In 2025, Google proposed one such approach based on a specific zero-knowledge proof system (Google, 2025). However, their approach relies on very recent advances in cryptography, is highly complex, and has yet to be verified by the community.

### 4.6.2 Offline Assurance – Credential-Based: Analysis

Pro:

- *Verification can be assured.* Service providers can be sure that age has been officially verified. They do not need to rely on the correctness of the information provided by the device.
- *Strong privacy protection.* When implemented with appropriate unlinkable credentials and selective disclosure this approach offers strong privacy protections.

Cons:

- *Non-private credential approaches exist.* The strong privacy protections disappear when linkable credentials are used. This is for example the case when using batch-issued credentials.
- *Technical Complexity.* Noticeably more technically complex, requiring significant work towards integrating this for service providers, and the development of wallets that support these credentials on users' devices.

## 4.7 Summary and Preliminary Conclusions

We make the following observations and recommendations (see also Table 4.3). We focus here on the technical analysis of the architectures. See Section 3 for the analysis of the different age assurance approaches.

- *Online Assurance.* Due to the high risks of tracking and profiling in the online assurance architecture, we do not recommend it. The only exception is the case of age inference where the AAS/IdP is the same as the SP. In this case, the online setting does not necessarily introduce additional privacy risks.
- *Offline Assurance – Device-based.* The device-based architecture has the advantage that it is very easy to implement for service providers, it just requires a (limited) one-time effort by OS manufacturers. It also does not create new privacy risks. Due to the limited compatibility with age inference, we do not recommend it. Finally, a combination with parental control and consent is possible as a combination.
- *Offline Assurance – Credential-based.* The credential-based architecture adds technical complexity, but in return is a little harder to circumvent. We recommend combining it with age verification or age estimation. It also works when social media platforms (or others that have a lot of data about users) act as AAS/Idp. Combining credentials with parental controls does not make sense.

|  | Age Verification | Age Estimation | Age Inference | Parental Control |
|---|---|---|---|---|
| Online Assurance | Not recommended | Not recommended | Ok (when SP performs inference) | Not possible |
| Offline Assurance (Device Based) | Recommended | Recommended (when on device) | Not recommended | Recommended as combination |
| Offline Assurance (Credential Based) | Recommended | Recommended | Ok (if platform acts as issuer) | Not applicable |

**Table 4.3: Recommended combinations of approaches (rows) and architectures (columns)**

# 5 General Limitations and Side Effects

Now that we have discussed in detail different approaches to age assurance and architectures in which these can be used, we zoom out and look at generic limitations and side effects of AAT.

## 5.1 Limitations Regarding Effectiveness

All age assurance technologies should ideally be effective in preventing minors from accessing the harmful material. After all, that is the whole point of introducing these technologies. Yet, essentially all of the above approaches are vulnerable to technological circumvention. These circumvention techniques are generic, and unlikely to go away any time soon without making very undesirable choices that limit the general availability of technology (see Side effects, 5.2.4).

Given that these circumvention technologies are likely to persist, they effectively define an *upper bound on the effectiveness* that any age assurance approach can provide. Age assurance can make it *harder* to access age restricted content, but definitely not *impossible*, in fact, far from impossible. The following tools and issues are of particular importance in limiting the effectiveness of AAT.

1. **VPNs and others.** Platforms are likely to only apply age assurance measures when they are forced to by law. While different jurisdictions take different measures, and some of these jurisdictions do not enforce age assurance, age assurance can be circumvented through digital technologies. One key example is *Virtual Private Networks* (VPNs), which enable users to pretend that their device is connecting from another country. By choosing their country carefully, access to restricted content can easily be enabled.
   *Impact on Age Assurance.* Circumvention through VPNs does not affect all instances and cases of age assurance to the same degree: a) This type of circumvention does not work when using *Parental Control and Consent* tools as these include measures to make it difficult to install or use VPN services. b) The effectiveness of this circumvention measure likely depends on the type of age-restricted content. It is much harder to move your Google or Apple account to another jurisdiction than it is to spin up a VPN connection to visit a porn website.
2. **Adult by Proxy.** By design, adults *are* allowed to access restricted content. Parents or other adults could therefore "enable" a child's device to effectively act like an adult's device.

*Impact on Age Assurance.* This approach is particularly effective in undermining measures of age restriction when verification happens only once or rarely: this includes all offline age assurance architectures, as well as any service provider that stores the result of an online age verification as part of a user's account.

It is important to note that the adult assisting in this process does not have to be the actual parent or legal guardian of the child. Any adult suffices. An analogy from the real-world for this type of circumvention would be a grown-up buying alcohol or cigarettes for an underaged person.

3. **Transfer of Devices/Accounts.** Related to the previous point, there is no age assurance approach that prevents an adult from setting up a device or account as their own, only to then let a child use it.
*Impact on Age Assurance.* This approach is particularly effective for offline architectures as well as any setting with user accounts. However, online architectures are also likely to store verification results for a while. The only approach that might detect a change of account or device owner is continuous age inference, where changed patterns may reveal a user change, prompting a request for re-assessment of age.

## 5.2 General Side Effects

The introduction of mandatory age assurance likely also has side effects beyond the direct consequences of technological choices. We discuss the most important ones here.

### 5.2.1 Privacy and Surveillance

Age assurance systems create the risk of enabling or forcing *increased tracking* of users to make the burden of age assurance manageable. This particularly applies to systems where (i) the service provider is responsible for assessing a user's age (i.e., online and credential-based architectures); and (ii) users do not normally have an account. Examples of settings where users typically do not have an account are pornographic websites, websites offering (also) age-restricted or violent content, etc.

When such providers have no technical means to easily recognize returning visitors, they must request age assessment again. Since age assurance is burdensome for the user – especially for online architectures, but to a reasonable extent also in credential-based architectures, this creates pressure to increase tracking to facilitate recognizing recurring users. Worse, when using private browsing modes (as you'd expect for some of these types of content) cookies cannot be used to store the results of age assurance. Hence, service providers might encourage users instead to create accounts, so that service providers can associate behavioral analyses and age-verification status with an account.

As a result, *the introduction of age assurance technologies has the risk of reducing the privacy of all users*. In settings where before accounts and cookies were not necessary, users could reduce the impact of tracking by refusing to create an account and to use private browsing modes. Yet, the introduction of age verification makes it much more likely that users' privacy is impacted. Both platforms and users are likely to prefer more persistent modes so that age verification does not need to happen repeatedly.

## 5.2.2 Discrimination of Content, Function Creep, and Potential for Censorship

Age assurance technologies are suggested with the intention to reduce harm and if they are effective, they indeed have the potential to significantly reduce harm by preventing minors from accessing harmful materials. However, at their core, they are tools to discriminate not only between user groups (adults versus minors), but also types of content (harmful versus harmless). As such, any age verification technology can be used for online censorship. This can happen in three ways.

First, censorship can occur by extending the definition of which websites (or apps, or resources) should be age restricted. These measures are likely to result in additional censorship. On the one hand, anyone below the age limit can clearly not access these sources anymore. On the other hand, any age restriction is likely to reduce the traffic to such websites by people that are of the right age, as is happening already for porn sites in the UK (BBC, 2025). It is easy to see how the argument could be made to extend age restrictions to other websites. We already see in other countries that topics such as reproductive health, gender identity, and certain books are subject to censorship under the heading of child protection. Similar arguments could easily be made to restrict access to information for teenagers. In fact, schools for example have already been found to block relevant information (TheMarkup, 2024).

Second, censorship occurs when age restrictions are applied too broadly to specific sources. Historically and internationally, such cases are not without precedent. Several countries have already applied blanket blocks of social media websites such as X/Twitter and Youtube because of the presence of undesirable content (Freedom

House, 2025). The online encyclopedia Wikipedia offers another example of a source that might be at risk of full age verification requirements. It is a large platform, has many users, and at least some of its pages contain content where the argument could be made that these should be age restricted. Does this mean that all of Wikipedia is subject to age verification? Or only specific pages?

Third, age verification tools can easily be used to prevent other user groups from accessing online resources. There are no technical means to restrict the exclusion criteria to age information only. Any digital tool that could be used to verify a person's age online could be repurposed to verify other information stored, such as, legal residence status, gender, nationality, etc.

### 5.2.3 Discrimination against People and the Risk of Exclusion

While Age Assurance Technologies are indeed built with the goal to discriminate between adults and minors, i.e. to exclude certain groups from certain types of content, this process can also have wider and unintended side effects of illegitimately excluding different groups of adult users. And indeed, this may affect most severely user groups which are already marginalized. There are several reasons for this:

- *Difficulty in dealing with new or complicated technology*. People that lack technological sophistication, including elderly people, might struggle to learn or adjust to having yet another required step when using technology that is already difficult for them.

- *Requiring special or expensive hardware.* Age verification techniques implicitly or explicitly require special hardware (e.g. personal devices with a camera so that selfies can be taken for age estimation; possession of a mobile phone; or a mobile phone with specific hardware). While they are legally allowed to access the web, social media apps, etc. the difficulty and cost of obtaining such hardware for some people might hamper the ease with which they can access these sources.

- *Requiring specific documents.* Age verification typically relies on having specific forms of documents, for example identity documents; or having preexisting relationships with organizations that rely on such documents. Yet, some groups are much less likely to have access to documents: children below a certain age, international visitors, homeless people, and migrants. Additionally, EU citizens legally residing or visiting another European country might have difficulty obtaining any assumed nation-specific ID, especially when age verification is rolled out at the national or local level, rather than at the EU or global level.

Some of these restrictions can be lifted by ensuring appropriate fallbacks are available, especially for some of the issues caused by requiring hardware or specific documents. For example a direct document verification might work with international identity documents; even for people that do not have access to the required national or digital identity documents. Or users might be asked for credit card information as a replacement for specific hardware.

However, even when such fallbacks work, they risk creating a multi-tiered system. Where some people (e.g. German citizens) can easily access age restricted content at little risk to their privacy (e.g. by using a credential-based approach); while others (e.g. international visitors) can only do so at the cost of serious privacy reductions; and others might never be able to access such content at all.

In the U.S., for instance, an estimated 11% of adults and 60% of 15-19 year-olds do not have a driver's license or similar photo ID (Wagener, 2025). Such individuals – including disproportionate numbers of marginalized or lower-income people – could be unjustly denied access to online content they are legally entitled to, simply because they cannot produce a required ID (Forland, 2025).

### 5.2.4 Collateral Damage: Closing the Internet

The requirement for age verification also creates or increases the risk of a closed internet as a whole, especially if circumvention methods are themselves forbidden to ensure the effectiveness of child protection online.

For example:

- **Forbidding the use of VPNs.** As we discuss in Section 5.1, VPNs can often easily be used to circumvent age restrictions. It is therefore tempting to ban, or try to block VPNs to ensure age restrictions are enforced. Already now, lawmakers are discussing banning VPNs in certain jurisdictions (EFF, 2025), which, by the way, are a common technical standard for employees connecting to their employer's IT infrastructure .

- **Forbidding the use of privacy technology such as Tor.** The Tor network can be used to anonymously browse the internet. As with VPNs, it can in many cases be used to side-step age restrictions. It might therefore be tempting to ban or block Tor users' connections.

- **Restricting OSes or applying OS-wide rules.** Another line of age-verification technologies we discuss performs age verification on the device. While legal measures might force large manufacturers and creators of operating systems such as Google, Apple, or Microsoft to integrate such checks into their mobile and desktop operating systems; it is less clear what this means for smaller producers and open source software. For example, a new law proposed in California would apply to all operating systems, including open source platforms such as Linux (PCGamer, 2026). Again, it might be tempting to disallow installing open source software or to run on open source hardware, in order to ensure that age verification is effective.

In all these cases, we argue that the consequences of forbidding or restricting access to these tools severely limit rights to information and to anonymity online and are thus highly undesirable.

# 6 Conclusions

The goal of this paper was to analyse different types of age assurance technologies to support informed decision making for children and youth protection in digital worlds. We therefore assessed the different types of approaches (age estimation, age inference, age verification, parental control and consent) on different architectures (online assurance, offline assurance – device based, offline assurance credential based) in regards to their a) effectiveness, b) side-effects, and c) acceptance.

With this matrix in mind, we aimed at finding the right level of detail regarding the technological underpinnings of the different approaches to be useful for ethical judgments and political decision making regarding age assurance technologies. Based upon our analyses, we have reached the following conclusions regarding age assurance technologies.

1. Age Assurance Technologies (AAT) are, by design, instruments of age-based discrimination: their entire purpose is to differentiate between individuals on the basis of age and to allow or deny access accordingly. This discriminatory function constitutes both the main objective of AAT and simultaneously its most significant source of systemic risks. AAT as a tool is thus highly ambivalent.
2. No AAT system has achieved, or is likely to achieve, perfect robustness against circumvention. Empirical evidence demonstrates that age verification mandates trigger surges in VPN adoption and other bypass strategies, particularly among technologically literate minors. We must accept that AAT will never ensure 100% effectiveness; demanding perfect efficacy would render all systems inadmissible, while ignoring the circumvention problem generates a false sense of security.

3. However, sometimes, online age-based differentiation may be politically or legally desirable and justified where it serves the protection of minors from content, interactions, or features that pose serious risks to their wellbeing and development.
4. In those cases, AATs are needed, but must be both effective and proportionate to the severity of the harm from which minors are to be protected. AAT must therefore both demonstrate adequate effectiveness in actually preventing access and have as few side effects as possible.
5. At worst, AAT may be both ineffective, if the very minors the system aims to protect can readily bypass it through technical means, while imposing significant privacy costs on the general population, thereby leaving side effects fully present while effectiveness is diminished.
6. As was shown, the three desirable properties of any AAT system (effectiveness, minimal side effects, societal acceptance) cannot be maximised simultaneously with currently available technologies. Enhancement along one dimension necessarily entails degradation along at least one other: currently, the most robust mechanisms impose bigger privacy costs or generate stronger circumvention incentives, and the most privacy-preserving solutions face adoption barriers.
7. Because circumvention is equally available against both intrusive and less intrusive mechanisms, slightly less reliable systems with substantially fewer side effects represent the sweet spot. We therefore recommend *device-based solutions and parental controls and consent approaches* as these are easy to implement, straightforward for service providers, and at least as robust as the effort required to configure a VPN. We want to stress nonetheless a significant caveat of this approach, namely that it concentrates additional market power in the hands of the providers of major mobile operating systems, namely Google and Apple.
8. Hierarchy of AAT mechanisms: Drawing on the analyses, this paper *recommends a graduated hierarchy of AAT mechanisms*, ordered by their balance effectiveness and of side effects. All recommended solutions are, to varying degrees, circumventable; however, their collateral harms are substantially more acceptable than those of maximally intrusive alternatives.

- ○ **First preference:** *Parental Control and Consent (PCC), on-device.* PCC has very low side effects but moderate effectiveness. It preserves the primacy of parental responsibility, operates entirely on-device without transmitting identity data to third parties, and is already widely deployed. Its principal limitations are that it works only where parents are engaged and digitally literate, and it functions most effectively on smartphones and tablets, and less so on desktops and gaming consoles.
- ○ **Second preference:** *Device-based solutions as a second line of defence.* These are rather effective with still few side effects. Device-based age estimation (using on-device biometric processing) or device-based verification of identity documents (via NFC reading of eIDs processed locally, for instance) keep all sensitive data within the device. The service provider receives only an age-threshold signal and not the underlying biometric or identity data.
- ○ **Third preference**: *Credential-based wallet solutions with strong privacy guarantees*. These are rather effective with reasonable side effects. Wallet-based solutions employing zero-knowledge proofs (ZKP) enable users to prove they meet an age threshold without revealing any other personal information. Importantly, credential-based wallets are the only mechanism that function reliably on desktop computers in addition to mobile devices. The EU Digital Identity Framework exemplifies the wallet approach. The parallel "mini wallet" currently being piloted in five Member States, though, does not guarantee unlinkability which would be required to preserve users' privacy.

9. Criminal Law: Where criminal law prohibits the provision of certain content to minors (prototypical example: online pornography), PCC alone is legally insufficient because it provides the service provider with no assurance that an age assessment has occurred. In such contexts, credential-based solutions are the minimum acceptable mechanism, as they enable the service provider to receive a proof of age-threshold compliance without collecting the user's identity or personal data.
10. Legal requirements for verifiable cryptographic properties and publicly audited, open-standard, peer-reviewed protocols: The AAT domain exposes a fundamental tension in EU regulatory philosophy: technology neutrality future-proofs regulation cannot guarantee the specific cryptographic properties (issuer and verifier unlinkability, and selective disclosure) upon which genuine privacy protection depends. This paper proposes a structured middle path: rather than mandating specific products, the legislature should mandate verifiable cryptographic properties and require conformity with publicly audited, open-standard, peer-reviewed protocols. For credential-based wallet solutions, this implies a preference for established, publicly developed schemes over proprietary, closed-source implementations whose privacy properties cannot be independently verified. The requirement for open-source, peer-reviewed cryptographic foundations is not a deviation from technology neutrality; it is its necessary complement.

11. Specific requirements instead of high-level principles: Legislators therefore should not stick to formulate high-level principles such as "data minimisation" or "privacy by design" when mandating AAT or defining their features, because such principles are sufficiently abstract to permit widely divergent implementations, i.e. some genuinely privacy-preserving, others privacy-invasive in practice while claiming nominal compliance. Implementing acts should specify, at a minimum: the requirements for issuer and verifier unlinkability as well as selective disclosure; on-device or user-controlled processing of biometric and identity data; the prohibition of data retention beyond the age attestation; and conformity with publicly audited, open-standard cryptographic protocols.
12. Prevent gatekeeping & market domination: Lastly, a word of caution: Major platform operators may risk exploiting AAT mandates to consolidate market power and expand data collection. Platforms that already possess verified identity data, like Google, Apple, or Meta, are positioned to offer themselves as trusted age attestation providers, becoming identity and compliance brokers that quietly (re-)centralise control. Furthermore, AAT compliance processes generate new data signals (age thresholds, verification timing, device information) that platforms may leverage for profiling and advertising. Legislators must ensure that AAT mandates do not inadvertently create new gatekeeper advantages that circumvent the objectives of the Digital Markets Act.
13. Information and Education: For people to make informed decisions about the use of AAT technologies, they need to know about their respective features and limitations. Introducing AAT policies and technologies should thus be accompanied by appropriate information campaigns.

We acknowledge that in balancing the three desiderata of children rights – protection, participation, and empowerment –, different conclusions may be drawn. Moreover, our analyses on the technical possibilities as well as the limitations and side effects of age assurance technologies should be complemented with insights from other domains of practice and disciplines, such as psychology, pedagogy, medicine and media studies, to inform political decision making with the best available evidence and assessments from different perspectives.

Puc et al. (2020). Andraž Puc, Vitomir Štruc, Klemen Grm. Analysis of Race and Gender Bias in Deep Age Estimation Models, EUSPICO 2020, https://new.eurasip.org/Proceedings/Eusipco/Eusipco2020/pdfs/0000830.pdf

Qin, Musetti & Omar (2023). Yao Qin, Alessandro Musetti, Bahiyah Omar. Flow Experience Is a Key Factor in the Likelihood of Adolescents' Problematic TikTok Use: The Moderating Role of Active Parental Mediation. Int J Environ Res Public Health. 2023 Jan 23;20(3):2089. https://doi.org/10.3390/ijerph20032089

Rodríguez-de-Dios (2018). Isabel Rodríguez-de-Dios, Johanna M.F van Oosten, Juan-José Igartua. A study of the relationship between parental mediation and adolescents' digital skills, online risks and online opportunities. Computers in Human Behavior (82), 186-198. https://doi.org/10.1016/j.chb.2018.01.012

Scimex (2025). Age verification trial results to enable kids' social media ban. 1 Sep 2025. https://www.scimex.org/newsfeed/expert-reaction-age-verification-trial-results-to-enable-kids-social-media-ban

Sevilla-Fernández et al. (2025). David Sevilla-Fernández, Adoración Díaz-López, Vanessa Caba-Machado, Juan Manuel Machimbarrena, Jessica Ortega-Barón, Joaquín González-Cabrera. Parental mediation and the use of social networks: A systematic review. PLoS One. 2025 Feb 3;20(2):e0312011. https://doi.org/10.1371/journal.pone.0312011

Sleigh (2023). Sophia Sleigh. Will Age Verification Stop Kids Accessing Porn Online? Most Brits Think Yes, Shows Poll. https://ca.news.yahoo.com/public-backs-age-verification-stop-060009353.html

SRC (2024). The Social Research Centre. Age Assurance Consumer Research. December 2024 https://www.infrastructure.gov.au/sites/default/files/documents/attachment-a-age-assurance-consumer-research-analytical-report-december-2024.pdf

Storyboard18 (2026). Meta study found parental controls had little effect on teens' compulsive social media use. https://www.storyboard18.com/digital/meta-study-found-parental-controls-had-little-effect-on-teens-compulsive-social-media-use-89933.htm

Teale (2024). Chris Teale. Americans are skeptical of online age verification, even as its use grows abroad. 11 Mar 2024. https://www.route-fifty.com/emerging-tech/2024/03/americans-are-skeptical-online-age-verification-even-its-use-grows-abroad/394798/

TechUK (2024). Julie Dawson. AI Adoption Case Study: learn about Yoti's facial age estimation tool! 14 Oct 2024 https://www.techuk.org/resource/ai-adoption-case-study-learn-about-yoti-s-facial-age-estimation-tool.html

TheMarkup (2024). Tara García Mathewson. Schools Were Just Supposed To Block Porn. Instead They Sabotaged Homework and Censored Suicide Prevention Sites. 13 Apr 2024. https://themarkup.org/digital-book-banning/2024/04/13/schools-were-just-supposed-to-block-porn-instead-they-sabotaged-homework-and-censored-suicide-prevention-sites

Trotman (2024). Rachael Trotman. Yoti facial age estimation evaluated in the NIST Face Analysis Technology Evaluation program.